\documentclass[%
 reprint,
 amsmath,amssymb,
 aps,
]{revtex4-2}

\usepackage{graphicx}
\usepackage{dcolumn}
\usepackage{bm}
\usepackage{siunitx}
\usepackage[version=4]{mhchem}
\usepackage{multirow}
\usepackage{tikz}
\usepackage{hyperref}

\begin{document}


\title{Electronic Structure of Cesium-based Photocathode Materials from Density Functional Theory: Performance of PBE, SCAN, and HSE06 functionals}

\author{Holger-Dietrich Sa{\ss}nick}
\affiliation{
Carl von Ossietzky Universit\"at Oldenburg, Physics Department, D-26129 Oldenburg, Germany
}
\author{Caterina Cocchi}%
\affiliation{
Carl von Ossietzky Universit\"at Oldenburg, Physics Department, D-26129 Oldenburg, Germany
}
\affiliation{
Humboldt-Universit\"at zu Berlin, Physics Department and IRIS Adlershof, D-12489 Berlin, Germany
}
\date{\today}

\begin{abstract}
The development of novel materials for vacuum electron sources in particle accelerators is an active field of research that can greatly benefit from the results of \textit{ab initio} calculations for the characterization of the electronic structure of target systems.
As state-of-the-art many-body perturbation theory calculations are too expensive for large-scale material screening, density functional theory offers the best compromise between accuracy and computational feasibility.
The quality of the obtained results, however, crucially depends on the choice of the exchange-correlation potential, $v_{xc}$.
To address this essential point, we systematically analyze the performance of three popular approximations of $v_{xc}$ (PBE, SCAN, and HSE06) on the structural and electronic properties of bulk \ce{Cs3Sb} and \ce{Cs2Te} as representative materials of Cs-based semiconductors employed in photocathode applications.
Among the adopted approximations, PBE shows expectedly the largest discrepancies from the target: the unit cell volume is overestimated compared to the experimental value, while the band gap is severely underestimated. On the other hand, both SCAN and HSE06 perform remarkably well in reproducing both structural and electronic properties. 
Spin-orbit coupling, which mainly impacts the valence region of both materials inducing a band splitting and, consequently, a band-gap reduction of the order of 0.2~eV, is equally captured by all functionals.
Our results indicate SCAN as the best trade-off between accuracy and computational costs, outperforming the considerably more expensive HSE06.

\end{abstract}

\maketitle

\section{Introduction}

Computational methods for material modelling have finally reached into almost all the areas of materials science and discovery, including those fields that, for historical reasons and scientific distance, are far away from condensed-matter physics.
Among them, the design and production of photocathode materials for particle accelerators is certainly worth a mention \cite{hern+08pt,dowe+08nimpra}.
The most recent advances in this field have imposed a paradigm shift driven by the necessity to develop vacuum electron sources producing increasingly focused particle beams with minimized mean transverse emittance~\cite{xian-teic15,smed+09aipcp,vecc+11apl}.
These characteristics can be best achieved through the development of photocathodes based on semiconducting materials which absorb visible light close to the infrared threshold.
Alkali antimonides and tellurides generally fulfill this requirement and have been extensively investigated in this context~\cite{doli+88,musu+18nimpra,xian-teic15,kark+14prl} along with conventional semiconductors like GaAs~\cite{orlo+04nimpra,kark+14prl,xu+16oc} and popular thermoelectric materials such as PbTe~\cite{pong72jap,li-schr17arxiv}. 

The search for novel semiconducting materials with optimized characteristics for photocathode applications greatly benefits from \textit{ab initio} quantum-mechanical simulations. 
Reliable predictions of material compositions and of the corresponding electronic properties are seen as a complement and, in some cases, even as a replacement of the expensive trial-and-error experimental growth and characterization procedures~\cite{dibo+97nimpra,baza+11apl,feng+17jap,ding+17jap,xie+17jpd,schm+18prab}.
To this end, multi-purpose databases of materials properties computed from density functional theory (DFT), such as NOMAD~\cite{NOMAD}, Materials Project~\cite{MaterialsProject}, the Open Quantum Materials Database~\cite{saal+13jom}, or Materials Cloud~\cite{MaterialsCloud}, are regarded with particular interest for a quick screening of suitable compounds and also as a source of input data for photoemission models~\cite{anto+20prb}.
Yet, the vast majority of the electronic-structure information stored in these databases are obtained under semi-local approximations of the exchange-correlation potential ($v_{xc}$).
As such, these data are likely suitable only for a \textit{qualitative} assessment of properties that are relevant for photoemission.

Recent studies on Cs-based multi-alkali antimonide crystals based on many-body perturbation theory (MBPT) provide state-of-the-art references for the electronic and optical characteristics of these materials~\cite{cocc+18jpcm, cocc+19sr}. 
However, these methods are too expensive to be used for high-throughput screening, or in the simulations of surfaces and defected systems, which more closely reflect the intrinsic features of the real materials at significantly large computational costs and complexity.
A good compromise is given by advanced approximations of $v_{xc}$, which have undergone impressive progress in the last decade~\cite{burk12jcp,mard-head2017}.
Recent developments include new parametrizations in the class of meta-GGA functionals~\cite{tao-mo16prl, patr+19prb, bart-yate19jcp, furn+20jpcl}, approximations based on machine-learning methods~\cite{zhou+19jpcl, dick-fern20nc}, and improved descriptions of the long-range dispersion interaction~\cite{peng+16prx, cald+19jcp, chak+20jctc, herm-tkat20prl}.
Range-separated hybrid functionals and meta-GGA approximations are presently efficiently implemented in many DFT packages~\cite{levc+15,barn+17,leht+18} and, as such, can be routinely used in the \textit{ab initio} simulations of large systems.

To reliably take advantage of DFT calculations in the study of Cs-based photocathode materials, a systematic comparison between the performance of 
against reliable references is urgently needed.
Similar studies were performed either on specific families of materials such as lead chalcogenides (see, \textit{e.g.,} Ref.~\cite{skel+14prb}), or on new functionals~\cite{mo+17prb,jana+18jcp,jana+18jcp1}.
General benchmark studies on solids~\cite{zhan+18njp,borl+19jctc,tran+19jap} are also a valuable starting point for reference, but the multi-atomic composition of photocathode materials as well as their peculiar chemical properties demand dedicated analysis.
This is precisely the scope of this work.

Herein, we investigate the structural and electronic properties of cesium antimonide (\ce{Cs3Sb}) and cesium telluride (\ce{Cs2Te}), two semiconductors largely employed in the context of photocathode research, in the framework of DFT with different approximations of $v_{xc}$.
Having the practical application in mind, we settled to consider functionals that are implemented and ``ready to use'' in a large number of DFT programs, but also represent different levels of exchange-correlation approximations.
To this end, we choose three well-established functionals representative for increasing levels of sophistication in the treatment of the exchange-correlation potential, namely, the semi-local Perdew-Burke-Ernzerhof functional~\cite{perd+96prl} implementing the generalized gradient approximation (GGA), the strongly constrained and appropriately normed (SCAN) parameterization of the meta-GGA~\cite{sun+15prl}, and the Heyd-Scuseria-Ernzerhof range-separated hybrid functional HSE06~\cite{heyd+03jcp}.
These functionals are prominent representatives of their respective class of approximations (GGA, meta-GGA, and hybrids) that are widely applied to solids.
Among the three considered approximations for $v_{xc}$, meta-GGA is certainly the least mature one, as testified by the number of new developments~\cite{tao-mo16prl,verm-truh17jpcc,meji-tric18prb,asch-kuem19prr} and recent benchmark studies~\cite{mo+17prb,jana+18jcp,jana+18jcp1,borl+20npjcm} dedicated to it.
In this developing scenario, the choice of considering SCAN is partly driven by our own positive experience with this functional.

After the optimization of the lattice parameters, which we benchmark against experimental values, we compute band structure and density of states of the target systems. The resulting band-gap is contrasted with $GW_0$ results from MBPT.
Our results show that the SCAN functional is an excellent choice in terms of accuracy and computational efficiency and an overall improvement to the PBE functional.
We finally assess the performance of the three aforementioned functionals in describing spin-orbit coupling effects, which are known to be relevant in the investigated materials, as they are composed of relatively heavy atomic species.
We find that spin-orbit coupling, which has a relatively small effect on the electronic properties of both materials, is equally captured by all considered $v_{xc}$ approximations.

This paper is organized as follows.
In Section~\ref{comp_methods} the theoretical background of our calculations is reviewed and the relevant computational details are summarized.
The body of results is presented in Section~\ref{results}, including the analysis of structural and electronic properties as well as spin-orbit coupling.
Summary and conclusions are reported in Section~\ref{summary}.

\section{Theoretical Background and Computational Details}\label{comp_methods}

\begin{figure*}
\centering
\includegraphics[scale=1]{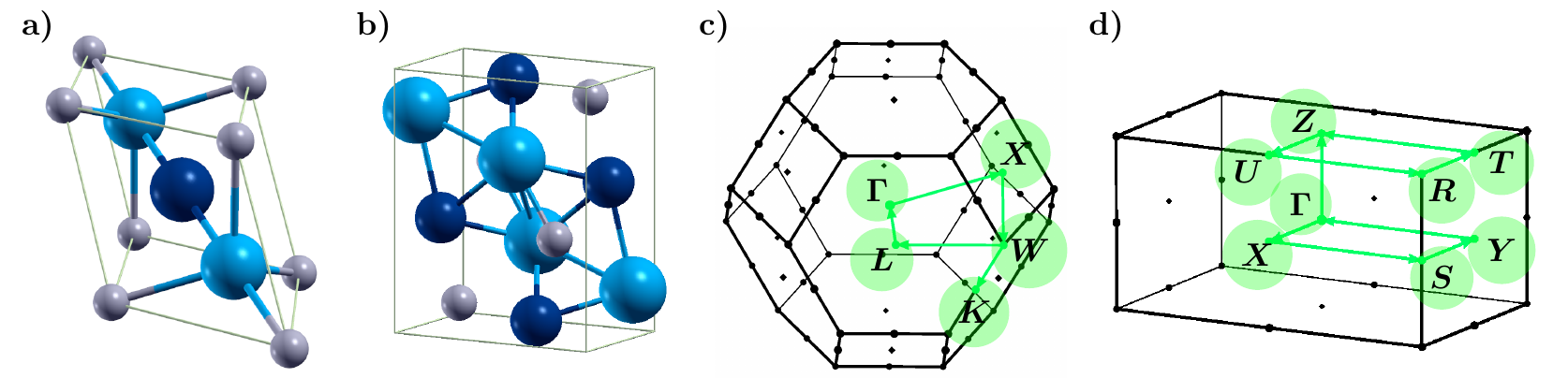}

\caption{\label{fig:unitcell} Ball-stick representation of the primitive unit cells of a) \ce{Cs3Sb} and b) \ce{Cs2Te}. Cs atoms in blue (light blue for Cs1 and dark blue for Cs2) and Sb and Te atoms in grey.
Brillouin zones of c) \ce{Cs3Sb} and d) \ce{Cs2Te} with high-symmetry points and the path connecting them highlighted.
Graphics created with XCrysDen~\cite{xcrysden}.}
\end{figure*}

The results of this work are obtained using DFT~\cite{hohe-kohn64pr} within the Kohn-Sham (KS) scheme, which maps the many-body system into an auxiliary system of non-interacting electrons with the same density as the fully interacting one~\cite{kohn-sham65pr}.
In this framework, one has to solve for each (valence) electron $i$ the KS equations, which in atomic units read:
\begin{equation}
\left[-\dfrac{1}{2}\nabla^2 + v_s(\mathbf{r}) \right] \varphi_i(\mathbf{r}) = \epsilon^{KS}_i \varphi_i(\mathbf{r}).
\label{eq:KS}
\end{equation}
The eigenvalue $\epsilon^{KS}_i$ is the KS energy per particle and the eigenfunction $\varphi_i(\mathbf{r})$ the KS wave function.
The Hamiltonian operator in square brackets contains two terms, namely the kinetic energy operator and the effective potential per particle
\begin{equation}
v_s(\mathbf{r}) = v_{ext}(\mathbf{r}) + v_H(\mathbf{r}) + v_{xc}(\mathbf{r}),
\label{eq:v_KS}
\end{equation}
including the external potential $v_{ext}$, accounting for the electron-nuclear interaction in absence of external fields, the Hartree potential, $v_H$, and, the exchange-correlation potential $v_{xc}$, the exact form of which is unknown.

The simplest approximation of $v_{xc}$, proposed already by Kohn and Sham in their seminal paper~\cite{kohn-sham65pr}, consists of treating exchange and correlation effects as in the homogeneous electron gas.
On the next rung, we find the GGA, in which $v_{xc}$ is derived from the local electron density and its gradient.
The PBE parameterization of GGA is adopted in this work. 
Taking one step further consists of approximating $v_{xc}$ including also the kinetic energy density: this is the so-called meta-GGA approximation.
As a representative of this level of approximation, which is currently explored in a number of development and benchmark studies ~\cite{tao-mo16prl,verm-truh17jpcc,meji-tric18prb,asch-kuem19prr,mo+17prb,jana+18jcp,jana+18jcp1,borl+20npjcm}, we employ herein the SCAN functional~\cite{sun+15prl}, which fulfills all the constraints for semi-local exchange-correlation functionals and has shown superiority in comparison to many other local and semi-local approximations on a broad range of systems~\cite{zhan+18, chen+17}.
Very recently, two regularized versions of the SCAN functional have been proposed, namely rSCAN~\cite{bart-yate19jcp} and r$^\mathrm{2}$SCAN~\cite{furn+20jpcl}.
The performance of the former is reported in the Appendix.

In parallel, another class of approximations for $v_{xc}$ has been developed in the last two decades, the so-called \textit{hybrid functionals} incorporating a fixed amount of Hartree-Fock exchange correcting the underlying semi-local DFT approximation~\cite{cson+10jctc}. 
Thanks to the presence of a fraction of Hartree-Fock exchange, the self-interaction error, which affects also meta-GGA functionals~\cite{zhan+18}, is alleviated.
More recently, this recipe has been further improved through the introduction of \textit{range-separated hybrid functionals}, in which the treatment of the exchange-interaction additionally depends on the range between the orbitals.
Here, as a representative of this class, we use the HSE06 functional~\cite{heyd+03jcp} which has shown remarkable success in the description of bulk properties for semiconducting and insulating condensed systems \cite{hend+11pssb}.
In this functional, the electron-electron interaction is split into a short-range and a long-range part: The short range part of the functional consists of 75$\%$ of the semi-local PBE xc functional and 25$\%$ of non-local Hartree-Fock exchange; in the long range part the functional coincides with PBE.

Other relevant aspects in the solution of the KS equation concern the treatment of core electrons and the choice of the basis set.  
In this work, we choose the all-electron full-potential implementation of FHI-aims~\cite{aims}
This package uses numerical atom-centered orbitals as basis set for all electrons.
The zero-order regular approximation is applied to correct for relativistic effects of the core-electrons.
Additionally, a non-self-consistent post-scf approximation is applied to account for spin-orbit coupling~\cite{aims_soc}.
These calculations are carried out with \textit{intermediate} settings for basis-set and integration grids.
A 18$\times$18$\times$18 and a 6$\times$12$\times$6  \textbf{k}-grid is chosen for \ce{Cs3Sb} and \ce{Cs2Te}, respectively.
In both cases the convergence of the \textbf{k}-grid is checked for the PBE functional.
The unit cells of both materials are optimized separately for each functional by calculating the stress tensor until the forces acting on each atom are converged below \SI{1.0e-3}{\electronvolt\per\angstrom}.

In absence of reliable experimental references for \ce{Cs2Te} and \ce{Cs3Sb} single crystals, we benchmark the accuracy of the DFT band gaps against the outcomes of $GW$ calculations~\cite{hedi65pr} in the partially self-consistent flavor $GW_0$~\cite{vonb-holm96prb,shis-kres07prb}.
In this formalism, the screened Coulomb potential $W_0$ is evaluated in a single-shot procedure, while the single-particle Green's function $G$ is updated self-consistently. 
These calculations are performed with the all-electron full-potential code \texttt{exciting}, which implements the linearized augmented plane wave plus local orbital method~\cite{exciting,nabo+16prb}.
The size of the basis set is determined by the radius of the muffin-tin (MT) spheres around the nuclei and the plane wave cutoff. 
The local orbitals available in the default species files are included in the calculations.
We anticipate that this can lead to a systematic underestimation of the band gaps~\cite{nabo+16prb}.
A MT radius of 1.8 and 2.5~bohr is used for both species in \ce{Cs2Te} and \ce{Cs3Sb}, respectively.  
A plane-wave cutoff of 8.0~Ha is adopted in both cases.
$GW_0$ calculations are performed on top of the PBE electronic structure obtained upon sampling the Brillouin zone with a 8$\times$8$\times$8 \textbf{k}-mesh in \ce{Cs3Sb} and 4$\times$8$\times$4 in \ce{Cs2Te}.
In the $GW_0$ step, a 6$\times$6$\times$6 \textbf{k}-grid is used for \ce{Cs3Sb} and a 3$\times$6$\times$3 one in \ce{Cs2Te}. 
Screening is computed in the random-phase approximation including 200 empty states in both systems.
It is worth noting that, while the self-consistency on the $G$ alleviates the starting-point dependence of the $GW_0$ calculations, a dependence of the results on the underlying approximation for $v_{xc}$ (PBE, in this case) cannot be entirely ruled out. 

\section{Results}\label{results}

\subsection{Structural Properties}\label{crystal}

We begin our analysis by inspecting the structural properties of both compounds.
Two different crystal structures have been reported for \ce{Cs3Sb}: a partially ordered \ce{NaTl} structure~\cite{kenn+57}, and a completely ordered anti-\ce{BiF3} structure~\cite{gnut+61}.
Here, we consider the latter geometry (space group $Fm\bar{3}m$), in line with previous theoretical studies~\cite{wei-zung87prb,kala+10jpcs, kala+10jpcs2,cocc+18jpcm,cocc20pssrrl}.
The primitive unit cell consists of three lattice sites for Cs atoms and one lattice site for Sb, see Figure \ref{fig:unitcell}a).
There are two crystallographically inequivalent lattice sites for Cs, characterized by different coordination and bonding character, which here are denoted as Cs1 and Cs2.
The fractional coordinates of Cs1 are $(\frac{1}{4}, \frac{1}{4}, \frac{1}{4})$ and $(\frac{3}{4}, \frac{3}{4}, \frac{3}{4})$ while Cs2 is located at $(\frac{1}{2}, \frac{1}{2}, \frac{1}{2})$.
The Sb atom is at the origin, $(0,0,0)$.

The experimental lattice parameter of \ce{Cs3Sb}, as obtained using x-ray and electron diffraction methods~\cite{kenn+57, gnut+61, robb-beck73jpd, baro+89}, ranges from 9.14~\AA{} to 9.19~\AA{}.
In Table~\ref{tab:lattice_Cs3Sb}, we compare our DFT results with the above-cited experimental references and, unsurprisingly, we find that PBE overestimates the lattice parameter by 1.18$\%$.  
This is a well-known problem of this functional and has also been reported for many other bulk materials \cite{schi+11jcp,tran+16jcp,zhan+18njp}
Also HSE06 tends to overestimate the lattice parameter, which is understandable considering that the fraction of exact exchange is included on top of a semi-local approximation for $v_{xc}$.
However, this functional leads to a sizeable improvement over PBE, yielding lattice vectors only 0.33$\%$ larger than the experimental one.
The best agreement with experiment is obtained with the SCAN functional, which delivers a value coinciding with the upper boundary of the experimental range (see Table~\ref{tab:lattice_Cs3Sb}).
The computed unit cell volumes reflect this behavior.
A very similar trend in accuracy for the three used functionals has been observed by Zhang et al. in the analysis of 64 different bulk solid materials~\cite{zhan+18njp}.

Turning to the cohesive energy, which is calculated by subtracting the total energy of free atoms from the total energy of the crystal unit cell, the SCAN and the HSE06 functionals yield very similar results, with the former featuring a slightly more negative value (see Table~\ref{tab:lattice_Cs3Sb}).
This finding is consistent with two recent studies that agree on the superiority of the SCAN functional over PBE in the prediction of accurate formation energies~\cite{zhan+18, yang+19prb}.
The PBE functional underestimates the cohesive energy by 0.3--0.4~eV/atom.

\begin{table}
\caption{Lattice parameters of the conventional unit cell, volume per atom ($\Omega$), and cohesive energy per atom ($E_{coh}$) of \ce{Cs3Sb} and \ce{Cs2Te}.\label{tab:lattice_Cs3Sb}}
\begin{ruledtabular}
\begin{tabular}{lcccc}
\multicolumn{5}{c}{\ce{Cs3Sb}} \\
                     & PBE         & SCAN   & HSE06    & exp.  \\
\colrule
$a$ [\si{\angstrom}]                                          &  9.298 & 9.194     & 9.220 &  9.14--9.19\footnote{\cite{kenn+57, gnut+61, robb-beck73jpd, baro+89}} \\ 
$\Omega$ [\si{\cubic\angstrom}/atom] & 50.24    & 48.58   & 48.99  & 47.72--48.51 \\
$E_{coh}$. [eV/atom]                                    & -2.037  & -2.436 & -2.316 & -- \\
\colrule
  \multicolumn{5}{c}{\ce{Cs2Te}} \\
                     & PBE         & SCAN   & HSE06    & exp.  \\
\colrule
 $a$ [\si{\angstrom}]                                            &   9.542 & 9.250  & 9.481          &  9.109\footnote{\cite{sche-boet91}}                 \\
 $b$ [\si{\angstrom}]                                           &  5.845  & 5.845  & 5.811           & 5.871\footnotemark[2]                        \\
 $c$ [\si{\angstrom}]                                        &  11.591 & 11.598 & 11.557         &    11.494\footnotemark[2]               \\
 $\Omega$  [\si{\cubic\angstrom}/atom]         & 53.88      & 52.26      & 53.06   & 51.22 \\
$E_{coh}$ [eV/atom]                                    & -2.655    & -3.081    &  -3.037 & --                                    \\
\end{tabular}
\end{ruledtabular}
\end{table}

Two orthorhombic crystal structures have been identified for \ce{Cs2Te}: one with space-group symmetry $P2_12_12_1$~\cite{prin-cord84}, and, more recently, an anti-\ce{PbCl2} crystal structure belonging to the space group $Pnma$~\cite{sche-boet91}.
Herein, we focus on the latter.
The orthorhombic unit cell of \ce{Cs2Te} is characterized by the three lattice vectors $a$, $b$, and $c$, and, in addition, by six internal parameters $x_{1,2,3}$ and $z_{1,2,3}$, which are needed to define the atomic positions of the 8 cesium and 4 tellurium atoms.
The fractional coordinates of the four lattice sites are $(x_{1,2,3}, \frac{1}{4}, z_{1,2,3})$, $(\frac{1}{2} + x_{1,2,3}, \frac{1}{4}, \frac{1}{2} - z_{1,2,3})$, $(\frac{1}{2} - x_{1,2,3}, \frac{3}{4}, \frac{1}{2} + z_{1,2,3})$ and , $(1 - x_{1,2,3}, \frac{2}{4}, 1 - z_{1,2,3})$  with the integers $1,2,3$ representing the positions of Cs1, Cs2, and \ce{Te}, respectively.
The experimental lattice parameters of \ce{Cs2Te}, as obtained from x-ray diffraction measurements~\cite{sche-boet91}, are reported in Table~\ref{tab:lattice_Cs3Sb}. 
The internal parameters are given in the Appendix (Table~\ref{tab:internal}).

In the case of \ce{Cs2Te}, all the considered functionals overall overestimate the unit cell volume with respect to the measurements.
However, the same trend discussed for \ce{Cs3Sb} is obtained also for this material: SCAN yields results in closest agreement to the experimental ones, followed by HSE06 and, lastly, by PBE, which gives rise to the largest overestimation. 
The individual lattice vectors, $a$, $b$, and $c$ exhibit a slightly different behavior compared to the whole volume, computed as their product.
While the $a$ parameter is, like the volume, overestimated by all adopted $v_{xc}$ approximations, with SCAN offering the best performance  (+1.55$\%$ compared to the experiment) and PBE the worst one (+4.75$\%$), the $b$ vector is \textit{underestimated} by all functionals, with PBE and SCAN giving the same and best result in this case (-0.44$\%$ with respect to the experimental reference); the estimate of HSE06 is anyway satisfactory (-1.02$\%$ below the experimental target).
Finally, for the $c$ parameter, HSE06 yields the closest value in comparison with the measurement (+0.55~$\%$) while SCAN overestimates the target by +0.90~$\%$, performing slightly worse than PBE (+0.84~$\%$).
As for the cohesive energy, the SCAN and the HSE06 functionals give very similar values, while, once again, the PBE functional predicts the shallowest value, that is about 0.4~eV/atom more positive than the results obtained with the other functionals.

\subsection{Electronic Properties}\label{electronic_properties}

\begin{figure*}
\centering
\includegraphics[scale=1]{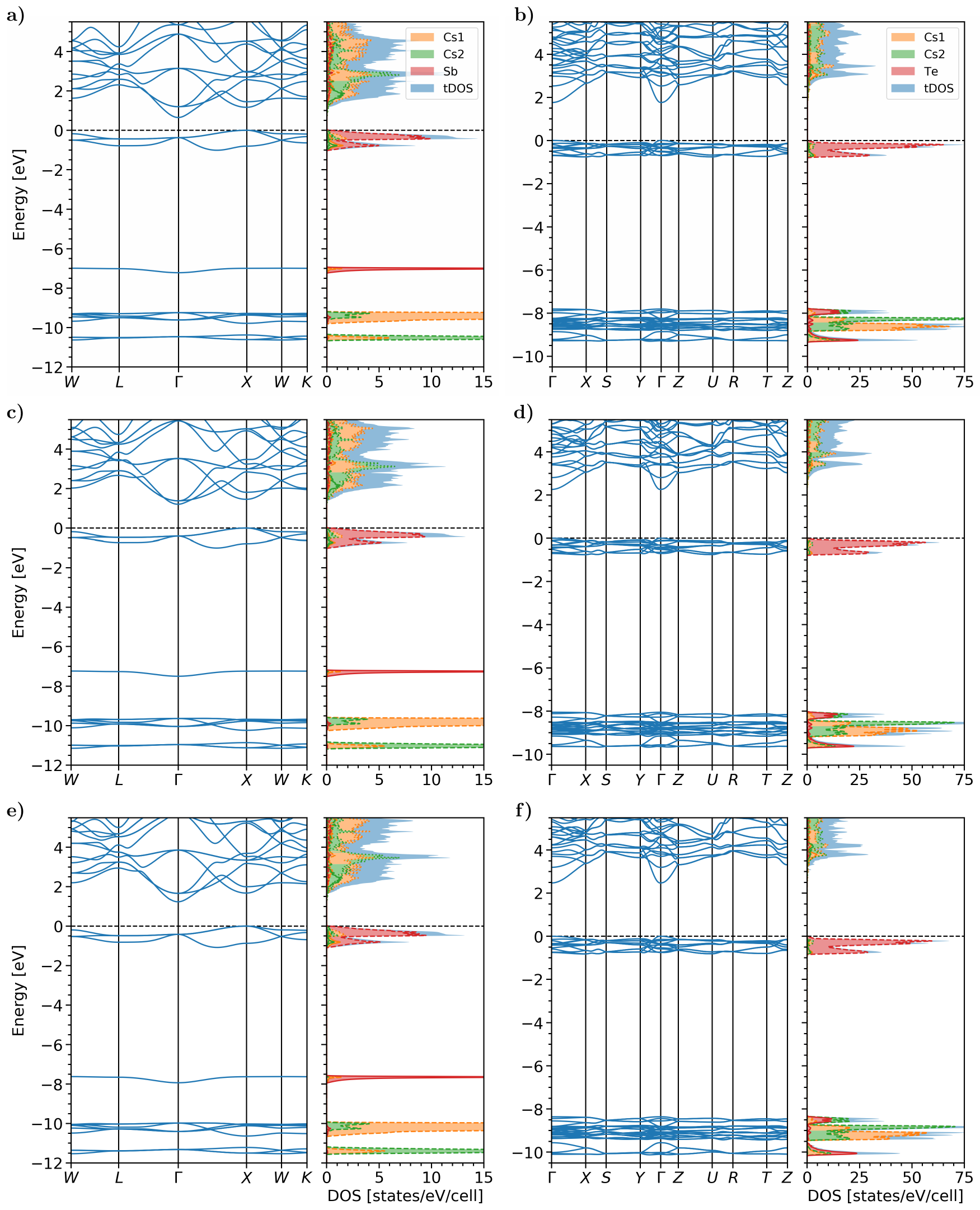}

\caption{\label{fig:bands_dos}Band structure and projected density of states  of \ce{Cs3Sb} (left) and \ce{Cs2Te} (right) obtained with PBE (panels a and b), SCAN (c and d), and HSE06 functional (e and f).
The energy is set to zero at the valence band maximum (dashed line).
 Solid, dashed, dotted and dash-dotted lines represent contributions from $s$-, $p$-, $d$- and $f$-orbitals, respectively.}
\end{figure*}

After analyzing the structural properties and stability of the considered materials, we now turn to their electronic structure, which is key to understand their behavior as photocathode materials.
We start by inspecting the band structure and the density of states (DOS), which is further decomposed into atom-projected contributions (see Figure~\ref{fig:bands_dos}). 

Bulk \ce{Cs3Sb} is characterized by an indirect band gap between the high-symmetry points $X$ and $\Gamma$, in agreement with the predictions of earlier studies~\cite{wei-zung87prb, kala+10jpcs,guo14,cocc+19sr}.
The highest occupied states of \ce{Cs3Sb} are dominated by \ce{Sb} electrons in the $5p$-shell, giving rise to three low-dispersive bands (see Figure~\ref{fig:bands_dos}, left panels).
The next valence state is 7.0--7.5~eV deeper in energy compared to the valence-band maximum, and is dominated by \ce{Sb} $s$-orbitals, as visualized in the projected DOS (PDOS).
Occupied \ce{Cs} states appear at even lower energies, around 9.5--10.0~eV (\ce{Cs1}) and 10.5--11.5~eV (\ce{Cs2}), also giving rise to flat ``atomic-like'' bands.
Identifying the contributions of chemically inequivalent atoms is particularly relevant in view of x-ray spectroscopy studies~\cite{cocc+16prb,vorw+17prb,cocc+19sr,cocc20pssrrl}.
In the conduction region of \ce{Cs3Sb}, bands are overall much more dispersive (see Figure~\ref{fig:bands_dos}, left panels).
The bottom of the conduction band at $\Gamma$ corresponds to the minimum of a parabolic state with \ce{Cs1}-\ce{Cs2} hybridized $sp$-contributions.
At higher energies, hybridized $s$- and $d$-states appear.

The general features discussed above, concerning band character and dispersion, are equally reproduced in all our calculations, regardless of the approximation adopted for $v_{xc}$.
The quantity that is mainly affected by the choice of the xc functional is the size of the band gaps (see Table~\ref{tab:gaps}).
In the absence of reliable experimenal references (it is very challenging to grow and characterize single crystalline \ce{Cs3Sb} samples~\cite{cult+16jvstb,gald+20jcp}), we benchmark our
DFT results  against those obtained with the $GW_0$ approximation of MBPT.
A comparison of the band structure between DFT (PBE functional) and the single-shot $G_0W_0$ approximation has been reported  for \ce{Cs3Sb} in Ref. \cite{cocc+19sr}.
As discrepancies concerning band shape and dispersion are found to be marginal, herein, we solely focus on the absolute values of the band gaps.

Given the indirect nature of the electronic gap in \ce{Cs3Sb}, which is correctly reproduced by all three considered functionals, it is instructive to inspect also the smallest direct band gap at $\Gamma$, hereafter addressed as \textit{optical gap}.
Unsurprisingly, the PBE functional underestimates the electronic gap by almost 50\% compared to the $GW_0$ method, in line with previous studies~\cite{wei-zung87prb, kala+10jpcs,guo14,cocc+19sr}.
On the other hand, band-gap values in very good agreement with the benchmark reference are obtained with both SCAN and HSE06: differences with respect to the $GW_0$ gaps are on the order of 15~meV.

Similar trends are obtained also for the optical gaps, however, with some remarkable differences (see Table~\ref{tab:gaps}).
The PBE result is only about 35$\%$ below the $GW_0$ one, in agreement with previous studies~\cite{wei-zung87prb, kala+10jpcs, cocc+19sr}, while both SCAN and HSE06 values depart from the reference by about 100~meV.
The reason for the different behavior of optical and electronic gaps can be understood through a close inspection of the band structures (see Fig.~\ref{fig:bands_dos}, left panels).
In the conduction region, the relative energy separation between the lowest-lying bands is critically sensitive to the xc functional.
While PBE features a separation of about 0.54~eV between the conduction band minimum and the next bands around $\Gamma$ [see Fig.~\ref{fig:bands_dos}a)], this value is drastically reduced in the band structures obtained with SCAN  [0.17~eV, see Fig.~\ref{fig:bands_dos}c)] and also with HSE06 [Fig.~\ref{fig:bands_dos}e)], even though to a slightly lesser extent (0.43~eV).
This variation results in an overall reduction in the energy difference between the conduction band minimum at $\Gamma$ and the subsequent minimum at $X$ of 267~meV for SCAN and 74~meV for HSE06.
At higher energies the unoccupied states are reproduced analogously by all three functionals, although absolute energies are generally lower in the PBE result compared to the other two. 
We see a similar trend also in the valence bands: The energy separation between highest-occupied bands and the next (flat) one depends on the adopted xc functional (see Fig.~\ref{fig:bands_dos}, left panels).
In the PBE band structure, this band appears at approximately 7.0~eV, while its energy decreases in the SCAN (7.2~eV) and HSE06 result (7.6~eV).
The same trend is followed qualitatively and quantitatively also by the deepest manifold of occupied bands displayed in Fig.~\ref{fig:bands_dos}, left panels.

\begin{table}
\caption{Electronic and optical gaps including spin-orbit coupling of \ce{Cs3Sb} and \ce{Cs2Te} in eV.}\label{tab:gaps}
\begin{center}
\begin{ruledtabular}
\begin{tabular}{lccccc}
   \multicolumn{6}{c}{\ce{Cs3Sb}}  \\
                                                & PBE    & SCAN   & HSE06 & $GW_0$    & exp.  \\
\colrule
$E_{gap}$                                   & 0.65  & 1.21 & 1.24  & 1.22 & \multirow{4}{*}{ 1.60\footnote{\cite{spic58pr}}} \\
$E_{gap}^{opt}$                    & 1.03   & 1.46 & 1.65 & 1.56 & \\
$E_{gap}$ + SOC                & 0.51  & 1.06 & 1.10 & --       & \\
$E_{gap}^{opt}$ + SOC & 0.85  & 1.31 & 1.48 & --       & \\
\colrule
  \multicolumn{6}{c}{\ce{Cs2Te}}  \\
                                                  &  PBE        & SCAN & HSE06    & $GW_0$ & exp.  \\
\colrule
$E_{gap}$                                    & 1.76      & 2.26  & 2.47 & 2.63 &\multirow{4}{*}{3.3\footnote{\cite{powe+73prb}}} \\
$E_{gap}^{opt}$                     & 1.76      & 2.26  & 2.47 & 2.63 & \\
$E_{gap}$ + SOC                   & 1.58      & 2.06  & 2.30 & --  & \\
$E_{gap}^{opt}$ + SOC & 1.58       & 2.06  & 2.30 & -- & \\
\end{tabular}
\end{ruledtabular}
\end{center}
\end{table}

\ce{Cs2Te} has a direct band gap at the $\Gamma$-point (see Fig.~\ref{fig:bands_dos}, right panels): hence, in this material, electronic and optical gaps coincide.
Similar to \ce{Cs3Sb}, also in \ce{Cs2Te} the uppermost occupied bands bear mainly $p$-character of the anion, which is here the \ce{Te} atoms.
However, in this case, a manifold of 12 bands is found at the top of the valence region since the unit cell contains 4 Te atoms.
Also, in analogy with \ce{Cs3Sb}, the next occupied states are largely separated in energy from the latter (7.8--8.4~eV below the valence band maximum).
In this region, bands are highly hybridized despite their low dispersion, exhibiting contributions from both \ce{Te} $s$- and \ce{Cs} $p$-states. 
The conduction bands of \ce{Cs2Te} are slightly more dispersive than the valence ones, especially in the vicinity of the minimum at $\Gamma$, where the lowest-unoccupied band exhibits a parabolic-like shape, as a fingerprint of the $s$-orbitals of both species dominating that state.
Energetically higher conduction bands show low dispersion and the signatures of hybridization between $d$- and $s$-states of \ce{Cs} and \ce{Te}.

The above-mentioned features are qualitatively captured by all three considered functionals, as in the case of \ce{Cs3Sb}.
Again and expectedly, PBE leads to a underestimation of the gap by about 33$\%$ in comparison to our $GW_0$ result (see Table~\ref{tab:gaps}).
Our PBE result is in good agreement with the value reported in Ref.~\cite{terd+12prb}.
Also SCAN and HSE06 give rise to smaller band-gaps compared to the $GW_0$ reference by 350~meV and 150~meV, respectively.

Differences in the relative band energies induced by different approximations of $v_{xc}$ are less pronounced in \ce{Cs2Te} compared to \ce{Cs3Sb} (see Fig.~\ref{fig:bands_dos}, right panels).
The separation between the conduction band minimum and the next unoccupied band at $\Gamma$ varies only by 40~meV between PBE and HSE06 and, again, to a larger amount for SCAN (210~meV).
The width of the highest-energy valence band manifold is equally captured by all xc functionals. 
Deeper bands are found starting from -7.8~eV (PBE), -8.0~eV (SCAN), and -8.4~eV (HSE06) with respect to the valence-band top set to zero in Fig.~\ref{fig:bands_dos}, right panels.

The comparison of the computed band gaps with the experimental ones is a delicate matter.
Reference values are available for \ce{Cs3Sb}~\cite{spic58pr} and \ce{Cs2Te}~\cite{powe+73prb}, which both correspond to rather old measurements.
Given the well-known experimental difficulties in controlling the composition and the quality of the samples of both materials \cite{gald+20jcp,prat+15prab}, it is wise to handle these data with care in a comparison with \textit{ab initio} results obtained for stoichiometric materials in ideal bulk unit cells.
Under these premises, we notice that the experimental gap available for \ce{Cs3Sb}, obtained from photoconductivity and photoabsorption measurements~\cite{spic58pr}, matches well with the optical band gap calculated with HSE06.
On the other hand, the experimental gap of \ce{Cs2Te} (3.3~eV resulting from photoemission spectroscopy~\cite{powe+73prb}) overestimates all calculated data, including the one from $GW_0$, by at least 0.5~eV.
Systematic first-principles studies and new measurements are needed for a final assessment of this point.

\subsection{The Impact of Spin-Orbit Coupling}\label{soc}

\begin{figure*}
\centering
\includegraphics[scale=1]{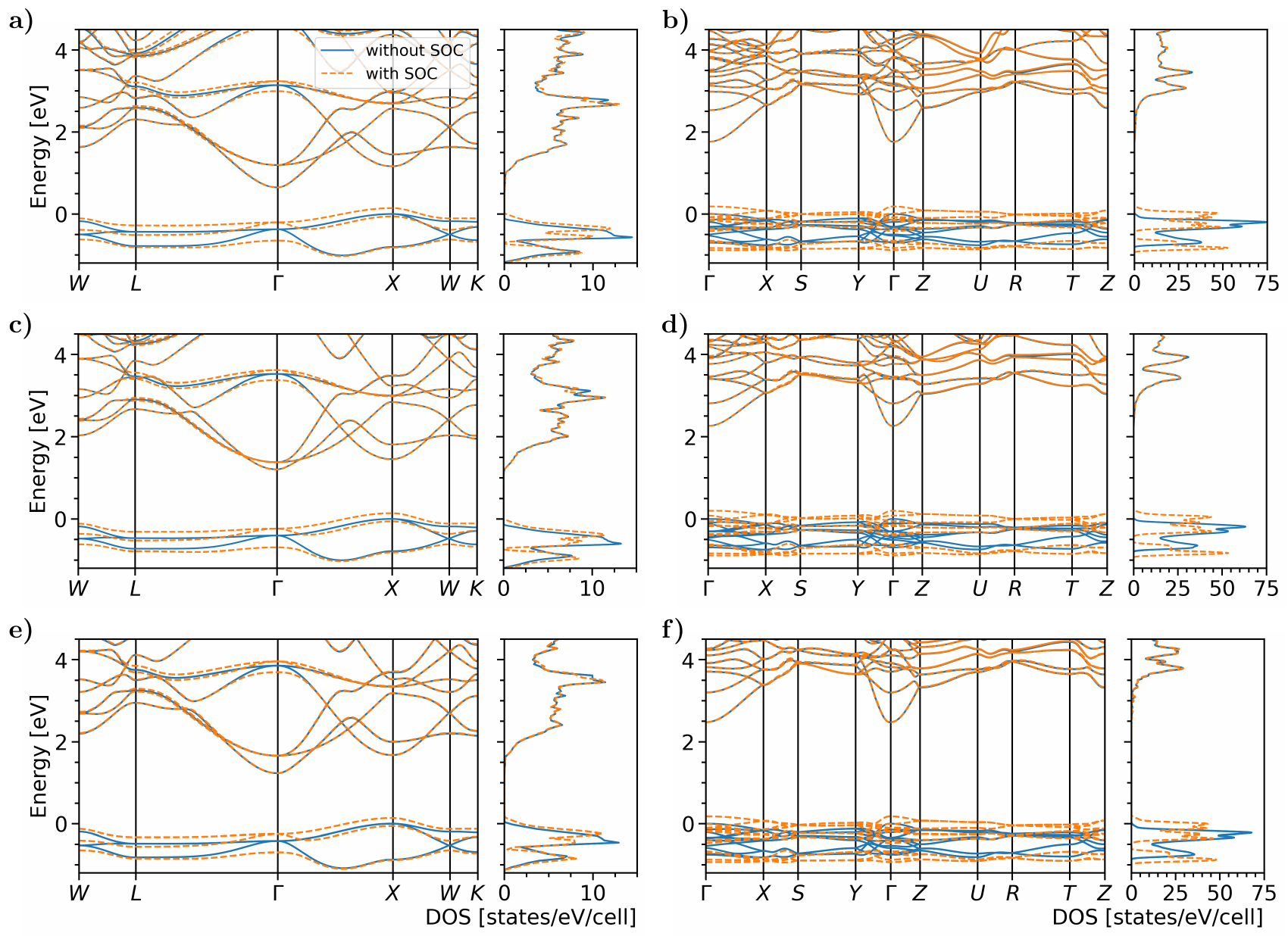}
\caption{Band structures and density of states close to the gap region of \ce{Cs3Sb} (left) and \ce{Cs2Te} (right), computed from DFT using a)-b) PBE, c)-d) SCAN, and e)-f) HSE06 functional with and without spin-orbit coupling (SOC), solid and dashed lines, respectively. The valence band maxima without SOC correction is set to zero in all plots.}\label{fig:soc}
\end{figure*}

Spin-orbit coupling (SOC) is a relativistic effect that can have a non-negligible impact on the electronic structure of materials formed by heavy atoms.
In order to assess the role of SOC in the two investigated systems,  \ce{Cs3Sb} and \ce{Cs2Te}, and the capability of the considered approximations for $v_{xc}$ to capture it, we performed an additional set of calculations, the result of which we contrast with those discussed above.
In this analysis, we focus on a window of approximately 5~eV around the gap region. 
At a first glance, it is clear that in both materials SOC affects the occupied bands much stronger than the unoccupied ones (see Fig.~\ref{fig:soc}).
In \ce{Cs3Sb}, the valence bands exhibit a SOC-induced splitting of about 0.25~eV and 0.45~eV at the high symmetry points $W$ and $\Gamma$, respectively.
Additionally, the uppermost occupied band is slightly shifted upwards under the effect of SOC.
Considering that the conduction-band minimum is unaltered by SOC, the band gap is overall decreased by approximately 0.15~eV, see Table~\ref{tab:gaps}.
A similar effect is noted also in the conduction region, again close to $\Gamma$, where the two degenerate bands at approximately 3~eV in Fig.~\ref{fig:soc}a) are split by 0.24~eV under the effect of SOC.
This feature is so localized in \textbf{k}-space that it is hardly appreciable in the plot of the DOS.

In \ce{Cs2Te}, the changes in the band structure induced by SOC are more pronounced than in \ce{Cs3Sb}.
The valence bands are pulled apart such that the upper states are up-shifted and the lower ones are down-shifted compared to their counterparts computed without SOC (see Fig.\ref{fig:soc}, right panels), resulting in an overall splitting of 0.2~eV.
As an additional effect, SOC reduces the band dispersion, especially along the path between the high symmetry points $S$ and $Y$ as well as between $Z$ and $U$, where the states are almost flat.
The conduction region of \ce{Cs2Te} is totally unaffected by SOC: both band structures and DOS displayed in Fig.\ref{fig:soc}, right panels, overlap with the results obtained without SOC.

No remarkable difference related to SOC splittings are appreciated among the results obtained with different approximations for $v_{xc}$ in both materials.
Differences in the calculated band-gaps are in line with those obtained neglecting SOC (see Table~\ref{tab:gaps}).
Our PBE and HSE06 results for \ce{Cs3Sb} follow qualitatively the same trend shown in Ref.~\cite{guo14}.
This is to be expected, as SOC is a physical effect that does not depend on the details of the approximations adopted for $v_{xc}$.
This finding has an important implication: If one is interested solely in the SOC splittings, a semi-local functional will be sufficient to capture them.   

\section{Summary and Conclusions}\label{summary}

We studied the performance of different levels of approximations for the exchange-correlation potential (PBE, SCAN, and HSE06) in DFT calculations on cesium antimonide and cesium telluride, both being actively used as photocathode materials for vacuum electron sources.
We focused on the structural and electronic properties of these two bulk crystals, including the effects of spin-orbit coupling. 
We found that all adopted approximations tend to overestimate the unit cell volume of both materials in comparison to the experimental references, with the PBE functional showing the largest discrepancies and the SCAN functional reproducing most accurately the experimental lattice parameters.
All explored functionals predict the same qualitative features in the electronic structures of the materials, including band-gap character, band dispersion, and atom-projected contributions to the total density of states.
Differences among the adopted approximations arise in terms of band energies, including the values of the electronic and optical gaps.
As expected, the PBE functional largely underestimates these quantities in both materials, while both SCAN and HSE06 yield values in much better agreement with the $GW_0$ reference, adopted in absence of reliable experimental data. 
The gaps resulting from the HSE06 approximation are the closest to the $GW_0$ benchmark, followed by those delivered by SCAN, which are systematically lower by about 0.2~eV.

In both materials, spin-orbit coupling induces band splittings in the valence region; in \ce{Cs3Sb}, additional splittings appear also in the conduction region. 
In \ce{Cs3Sb}, the band splittings occur in the vicinity of the high-symmetry points $\Gamma$ and $W$, while in \ce{Cs2Te} the internal splitting in the valence region its the width of this band by approximately 0.2~eV.
As a result of SOC, the band gap of both materials is decreased by a corresponding amount of energy.
These effects are reproduced almost identically by all considered functionals: approximations in the treatment of relativistic effects impact the description of this property much more strongly than the one on $v_{xc}$.

In conclusion, our results indicate SCAN as the optimal choice to approximate the exchange-correlation potential in DFT calculations of Cs-based antimonide and telluride materials.
Both structural and electronic properties are reliably reproduced at computational costs that are only slightly larger than those of generalized gradient approximations of $v_{xc}$ (see Table \ref{tab:comp_res} in the Appendix).
In agreement with the current knowledge, we found that PBE is suitable for simulating structural parameters but should be avoided in the quantitative description of the electronic properties.
Compared to the popular range-separated hybrid functional HSE06, in the considered materials, SCAN produces superior outcomes in terms of lattice vectors and unit-cell volumes.
On the other hand, HSE06 results for electronic and optical gaps are systematically closer to the $GW_0$ reference.
This is not surprising, as this hybrid functional has been specifically developed to accurately reproduce band gaps by means of DFT. 
The computational costs demanded by HSE06 are, however, generally higher compared to SCAN, which makes the latter a more convenient choice in the study of surfaces and defected bulks requiring large supercells. 
The ability of SCAN to reproduce the electronic properties of the considered Cs-based materials can be ascribed to the orbital character of the bands in the vicnity of the gap. 
As shown in Fig.~\ref{fig:bands_dos}, the valence and conduction regions of both \ce{Cs3Sb} and \ce{Cs2Te} are dominated by $s$- and $p$-orbitals, for which already the GGA approximation works fairly well. 
This finding and the proposed rationale suggests that any meta-GGA implementation may work well for these classes of systems. However, the numerical performance can critically depend on the details of the specific parameterization and should be therefore carefully assessed. 

The all-electron implementation adopted in this study makes our findings independent of the description of core levels via different pseudopotential recipes. 
The use of this approximation requires, therefore, dedicated benchmark studies.
Likewise, the choice of the basis set (\textit{e.g.}, plane waves) introduces additional parameters, such as the plane-wave cutoff, that have to be carefully evaluated for convergence.
Our work represents therefore a cornerstone for the systematic assessment of various approximations and parameters in the DFT simulations of crystalline photocathode compounds, and their impact in describing relevant properties. 
This is vital for predictive high-throughput screening on such systems, which can greatly contribute to the discovery and development of novel and more efficient materials for vacuum electron sources.

\begin{acknowledgments}
We are grateful for Jannis Krumland and Raymond Amador for their suggestions on the unpublished manuscript.
This work is funded by the German Federal Ministry of Education and Research (Professorinnenprogramm III) as well as from the Lower Saxony State (Professorinnen f\"ur Niedersachsen).
Computational resources provided by the North-German Supercomputing Alliance (HLRN), project bep00084, and by the HPC cluster CARL at the University of Oldenburg, funded by the DFG (project number INST 184/157-1 FUGG) and by the Ministry of Science and Culture of the Lower Saxony State.
\end{acknowledgments}

\section*{Appendix}\label{SI}

In Table~\ref{tab:internal}, we report the calculated internal lattice parameters of \ce{Cs2Te}.
In Table~\ref{tab:comp_res}, we compare the performance of PBE, SCAN, and rSCAN for calculations run on the same computational infrastructure.
It can be seen that both the average run time per self-consistent iteration ($t_{av}$) and the number of iterations needed for convergence are comparable.
The rSCAN functional shows a slightly improved performance  compared to the original SCAN functional.

\begin{table}[h!]
\caption{Internal lattice parameters for \ce{Cs2Te} computed from DFT with different approximations for $v_{xc}$ and compared with the available experimental reference~\cite{sche-boet91}.}
\begin{center}
\begin{ruledtabular}
\begin{tabular}{lcccc}
    & PBE         & SCAN   & HSE06    & exp. \cite{sche-boet91} \\
\hline
 $x_1$ & 0.028  & 0.035  & 0.029 & 0.016  \\
 $z_1$ & 0.172   & 0.183   & 0.173 & 0.180  \\
 $x_2$ & 0.146   & 0.154   & 0.147 & 0.152  \\
 $z_2$ & 0.572   & 0.571   & 0.572 & 0.568  \\
 $x_3$ & 0.246  & 0.241   & 0.246 & 0.248  \\
 $z_3$ & 0.886  & 0.886  & 0.885 & 0.888 \\
 \end{tabular}
 \end{ruledtabular}
\end{center}
\label{tab:internal}
\end{table}

\begin{figure}
\centering
\includegraphics[scale=1]{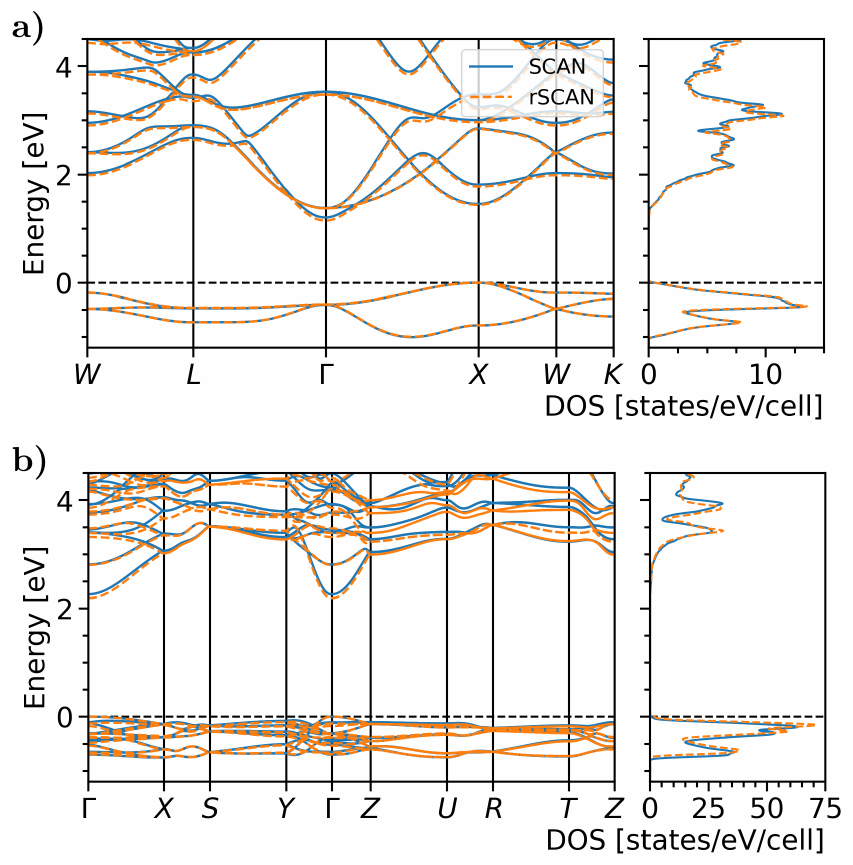}

\caption{\label{fig:comp_bands}Band structure and total density of states  of \ce{Cs3Sb} (top) and \ce{Cs2Te} (bottom) for the SCAN and rSCAN functional.}
\end{figure}

Tables~\ref{tab:comp_lattice} and~\ref{tab:comp_gaps}, as well as Fig.~\ref{fig:comp_bands} present the comparison between the performances of SCAN~\cite{sun+15prl} and rSCAN~\cite{bart-yate19jcp} functionals.
We notice slightly increased unit cell volumes and slightly more positive values of the cohesive energy given by the rSCAN functional in comparison with SCAN (Table~\ref{tab:comp_lattice}).
Differences in the band gaps are smaller than 0.1~eV, with the rSCAN functional yielding lower values (Table~\ref{tab:comp_gaps}).
However, no visible differences are found in the band structures computed with the two functionals (see Fig.~\ref{fig:comp_bands}.)

\begin{table}[h!]
\caption{Comparison among PBE, SCAN, and rSCAN functionals in terms of the average run time needed per self-consistent iteration ($t_{av}$), and the number of iterations needed to converge the solution of the Kohn-Sham equations.}
\begin{ruledtabular}
\begin{tabular}{ccccccc}
& \multicolumn{3}{c}{Cs$_3$Sb} & \multicolumn{3}{c}{Cs$_2$Te} \\
& PBE & SCAN & rSCAN & PBE & SCAN & rSCAN \\
\hline
$t_{av}$ [s]         & 4.43 & 5.49 & 5.36 & 7.71 & 10.31 & 10.07 \\
Nr. of iterations & 17     & 24     & 21      & 16    & 23 & 18 \\
\end{tabular}
\end{ruledtabular}
\label{tab:comp_res}
\end{table}

\begin{table}
\caption{Lattice parameters of the conventional unit cell, volume per atom ($\Omega$), and cohesive energy per atom ($E_{coh}$) of \ce{Cs3Sb} and \ce{Cs2Te} comparing the SCAN  and the rSCAN functional.\label{tab:comp_lattice}}
\begin{ruledtabular}
\begin{tabular}{lccccc}
\multicolumn{6}{c}{\ce{Cs3Sb}} \\
              & \multicolumn{3}{c}{$a$ [\si{\angstrom}]} & $\Omega$ [\si{\cubic\angstrom}/atom] &$E_{coh}$. [eV/atom] \\
              \colrule
SCAN  &  \multicolumn{3}{c}{9.194} & 48.58 & -2.436 \\
rSCAN &  \multicolumn{3}{c}{9.238} & 49.27 & -2.371 \\
\colrule
  \multicolumn{6}{c}{\ce{Cs2Te}} \\
 & $a$ [\si{\angstrom}] & $b$ [\si{\angstrom}] & $c$ [\si{\angstrom}] & $\Omega$  [\si{\cubic\angstrom}/atom] & $E_{coh}$ [eV/atom] \\
\colrule
SCAN   & 9.250 & 5.845 & 11.598 & 52.26 & -3.081 \\
rSCAN & 9.456 & 5.849 & 11.611  & 53.52 & -3.016 \\
\end{tabular}
\end{ruledtabular}
\end{table}

\begin{table}
\caption{Electronic and optical gaps including spin-orbit coupling of \ce{Cs3Sb} and \ce{Cs2Te} in eV comparing the SCAN and rSCAN functional.}\label{tab:comp_gaps}
\begin{center}
\begin{ruledtabular}
\begin{tabular}{lcccc}
   \multicolumn{5}{c}{\ce{Cs3Sb}}  \\
   & $E_{gap}$ & $E_{gap}^{opt}$ & $E_{gap}$ + SOC & $E_{gap}^{opt}$ + SOC \\
   \colrule
SCAN   & 1.21 & 1.46 & 1.06 & 1.31 \\
rSCAN & 1.15 & 1.44 & 1.01 & 1.30 \\
\colrule
  \multicolumn{5}{c}{\ce{Cs2Te}}  \\
   & $E_{gap}$ & $E_{gap}^{opt}$ & $E_{gap}$ + SOC & $E_{gap}^{opt}$ + SOC \\
\colrule
SCAN   & 2.26 & 2.26 & 2.06 & 2.06 \\
rSCAN & 2.19  & 2.19 & 2.00 & 2.00 \\
\end{tabular}
\end{ruledtabular}
\end{center}
\end{table}

\newpage


%

\end{document}